\documentclass[journal]{IEEEtran}

\usepackage{cite}
\usepackage{amsmath}
\usepackage{algorithmic}
\usepackage{url}
\usepackage{graphicx}
\usepackage{array}
\usepackage{amssymb}
\usepackage{multirow}
\usepackage[caption=false,font=footnotesize]{subfig}
\usepackage[pagebackref=true,breaklinks=true,colorlinks,bookmarks=false]{hyperref}
\usepackage{amsthm}
\usepackage{dsfont}
\usepackage{booktabs}
\usepackage{xcolor}
\usepackage[T1]{fontenc}
\usepackage{algorithmic}
\usepackage[ruled, vlined]{algorithm2e}
\usepackage{graphicx}
\usepackage{subfig}
\usepackage{booktabs}
\usepackage{threeparttable}

\SetKwInput{KwInput}{Input}             
\SetKwInput{KwOutput}{Output}
\graphicspath{{fig/}}

\def\al{\emph{et al. }}

\begin{document}
\title{Dual Dynamic Threshold Adjustment Strategy for Deep Metric Learning}

\author{
	Xiruo Jiang,
	Yazhou Yao,
	Sheng Liu,
	Fumin Shen,
    Liqiang Nie,
	and~Xian-Sheng Hua
	\thanks{X.~Jiang, and Y.~Yao are with the School of Computer Science and Engineering, Nanjing University of Science and Technology, Nanjing, China.}
	\thanks{S.~Liu is with the State Key Laboratory of Virtual Reality Technology and Systems, School of Computer Science and Engineering, Beihang University, Beijing, China.}
	\thanks{F.~Shen is with the School of Computer Science and Engineering, University of Electronic Science and Technology, Chengdu, China.}
	\thanks{L.~Nie is with the School of Computer Science and Technology, Harbin Institute of Technology (Shenzhen), Shenzhen, China.}
	\thanks{X.~Hua is with the Terminus Group, Beijing, China.}
}

\markboth{}%
{Shell \MakeLowercase{\textit{et al.}}: I2CRC}

\maketitle

\begin{abstract}

Loss functions and sample mining strategies are essential components in deep metric learning algorithms. However, the existing loss function or mining strategy often necessitate the incorporation of additional hyperparameters, notably the threshold, which defines whether the sample pair is informative. The threshold provides a stable numerical standard for determining whether to retain the pairs. It is a vital parameter to reduce the redundant sample pairs participating in training. Nonetheless, finding the optimal threshold can be a time-consuming endeavor, often requiring extensive grid searches. Because the threshold cannot be dynamically adjusted in the training stage, we should conduct plenty of repeated experiments to determine the threshold. Therefore, we introduce a novel approach for adjusting the thresholds associated with both the loss function and the sample mining strategy. We design a static Asymmetric Sample Mining Strategy (ASMS) and its dynamic version Adaptive Tolerance ASMS (AT-ASMS), tailored for sample mining methods. ASMS utilizes differentiated thresholds to address the problems (too few positive pairs and too many redundant negative pairs) caused by only applying a single threshold to filter samples. AT-ASMS can adaptively regulate the ratio of positive and negative pairs during training according to the ratio of the currently mined positive and negative pairs. This meta-learning-based threshold generation algorithm utilizes a single-step gradient descent to obtain new thresholds. We combine these two threshold adjustment algorithms to form the Dual Dynamic Threshold Adjustment Strategy (DDTAS). Experimental results show that our algorithm achieves competitive performance on CUB200, Cars196, and SOP datasets. Our codes are available at \url{https://github.com/NUST-Machine-Intelligence-Laboratory/DDTAS}.

\end{abstract}


\IEEEpeerreviewmaketitle

\section{Introduction}
Metric learning aims to either expand or reduce the distance between samples of different/same categories~\cite{jiang2020multi,hoi2010semi,cover1967nearest}. It is widely used in tasks like classification and detection~\cite{yao2017exploiting,jiang2022deep,yao2023automated}. With the development of GPU technology and deep learning~\cite{tang2023holistic,yao2021non,pei2022eccv,yao2021jo,sun2021webly}, deep metric learning has emerged as a prominent area of research. Compared with the traditional metric learning methods, deep metric learning transfers the tedious feature engineering to the deep neural network, simplifying the capture of nonlinear information through the nonlinear activation layer of the network. As a result, deep metric learning has demonstrated notable success in diverse areas including person re-identification~\cite{liu2023generative, zhang2017image,yang2018person}, image retrieval~\cite{zhong2017slmoml}, face detection~\cite{hu2014discriminative}, representation learning \cite{yao2019deep}, and few-shot learning~\cite{jiang2020multi,Authors28,Authors29}.

Deep metric learning algorithms can be divided into two categories, i.e., proxy-based and pair-based. ProxyNCA~\cite{Authors55} and ProxyAnchor~\cite{Authors63} are typical works based on the proxy algorithm. There are more algorithms based on pairs, such as contrastive loss~\cite{Authors16}, triplet loss~\cite{Authors21}, triplet-center loss~\cite{Authors37} and \textit{N}-pairs loss~\cite{Authors59}. In these pair-based algorithms, whether they rely on the absolute distances between samples or compute relative similarities based on anchor points, there is a common step of employing a threshold (margin) to select pairs with rich information. It is vital to choose an appropriate threshold since it will eliminate pairs with limited information and help accelerate the training convergence. However, there are existing shortcomings in selecting thresholds {for conventional algorithms}: (a) These thresholds are often obtained by manual adjustment. For familiar scenes, we can obtain thresholds based on previous experience. However, for new benchmarks, we need to spend expensive time and resources to determine the threshold; (b) The embedding distribution learned by the model varies during the training process. Initially, sample representations are dispersed in the embedding space. At this stage, if the threshold is too small to constrain the pairs, some pairs with insufficient information will enter the training step. When the threshold selection is too strict, some valuable sample pairs will be excluded from the training process.

Early deep metric learning loss functions utilize hinge functions to filter simple sample pairs directly (e.g., contrastive loss~\cite{Authors16}). Subsequent algorithms use this `0 and 1' sample mining strategy~\cite{Authors21, Authors7} as a module to filter pairs roughly. Some works, such as lifted structure loss~\cite{Authors49} and \textit{N}-pairs loss~\cite{Authors59}, design smooth functions to weight more informative samples. Wang \al~\cite{Authors7} summarized various weighting methods and used multiple similarities between positive and negative pairs to weight samples, achieving remarkable performance. However, these methods still rely on sample pairs that have passed the `static' threshold screening. Sample pairs with rich information are still likely to be excluded from the threshold range due to changes in the embedding space.

To alleviate this problem, Wu~\al~\cite{Authors44} proposed Distance Weighted Sampling (mining strategy) and Margin-Based Loss. Distance Weighted Sampling filters sample pairs based on the sample distance distribution. This work considers that the sample pairs should be drawn from the entire similarity region rather than the local region (Hard or Semi-Hard mining). By doing so, this method broadens the spectrum of selected pairs and can mine hard samples. However, in situations where batch sizes are restricted, this algorithm is insufficient for the constraint of pairs. Distance Weighted Sampling results in fewer precious sample pairs entering the loss than methods using semi-hard or hard sample mining strategies. At the same time, this strategy does not effectively deal with positive pairs. Another contribution of this work is to propose Margin-Based Loss. This loss function has a more flexible and optimized threshold $\beta$ to control the boundary of positive and negative pairs. However, this loss function still chooses the `hard' selection instead of a `soft' weighting schema to calculate the loss subsequent to the mining stage. 

Existing deep metric learning algorithms have non-negligible problems for selecting thresholds in sample mining and loss function design. Therefore, our work focuses on the design of thresholds in deep metric learning. We propose The Asymmetric Sample Mining Strategy (ASMS) for the mining strategy. Different from the existing pair selection algorithms that use the same threshold to filter positive and negative pairs, our ASMS uses a broader tolerance threshold for positive pair filtering, coupled with a more stringent threshold for negative pair screening. This algorithm tightens the selection of informative negative pairs while significantly increasing the number of positive pairs participating in the training process. To achieve a more informative threshold, we design a dynamic threshold adjustment strategy for both the loss function and sample mining strategy, respectively, and name this Dual Dynamic Threshold Adjustment Strategy (DDTAS).\\ 
In this paper, our contributions are as follows:

We propose the Asymmetric Sample Mining Strategy (ASMS) to solve the problem of an insufficient number of positive pairs participating in training, coupled with an excess of redundant negative pairs. ASMS uses different thresholds to filter positive and negative pairs separately. After using ASMS, the relatively loose positive pair threshold allows subsequent training steps to obtain more positive pairs. A more tight negative sample threshold can filter out more informative samples. Consequently, the remaining pairs furnish more valuable information, leading to an enhanced retrieval performance of the algorithm.

During training, the embedding space learned by the model is constantly changing. A tighter distribution of positive pairs and a more dispersed distribution of negative pairs reduces the effect of the initial mining threshold. Therefore, we dynamically adjust the mining threshold and re-mining the samples based on the current ratio of positive and negative pairs to be mined. This strategy is named Adaptive Tolerance ASMS (AT-ASMS).

We design a new metric learning loss function named Soft Contrastive Loss, which inherits the strengths of both Contrastive Loss~\cite{Authors16} and Binomial Deviance Loss~\cite{2014Deep}. 

The image retrieval performance of our algorithm is evaluated on three datasets CUB200~\cite{Authors46}, CARS196,~\cite{Authors47} and SOP~\cite{Authors49}. The experimental results show that our algorithm achieves competitive performance compared to the existing deep metric learning algorithm.

\section{Related Work}\label{sec2}
\subsection{Deep Metric Learning}\label{2.1}
In recent years, the development of deep convolutional neural networks and the update of powerful computing devices~\cite{schmidhuber2015deep,mao2023attention} have ushered in a new era for deep metric learning. These deep neural network models break through the limitations of linear algorithms' ability to distinguish complex data patterns. Moreover, when deep networks are employed alongside extensive sample data, deep metric learning algorithms effectively mitigate the overfitting typically encountered in nonlinear algorithms. Through this type of algorithm, we can transform the data from the original space to a more suitable embedding space. Better embedding space means better classification and clustering. One pioneering work proposed by Bromley \al is the Siamese Network~\cite{Authors14}. This network learns an embedding space where the features of the same class samples are close, and the samples of the different types are far away from each other. The versatility of the Siamese Network extends to applications in tasks like face verification and dimensionality reduction.

\textbf{Backbone Network.} There are few dedicated backbone networks for the backbone module for deep metric learning. The backbone networks currently applied to deep metric learning algorithms include GoogleNet, Inception series networks represented by BN-Inception~\cite{DL8_ioffe2015batch}, and ResNet50~\cite{DL7_he2016deep}. BN-Inception and ResNet50 are presently the most widely used backbone networks. According to the current work, for the same algorithm, using different networks will bring about different results. In this paper, we opt for BN-Inception as the backbone network.

\begin{figure*}[t]
	\begin{center}
		\includegraphics[width=1\linewidth]{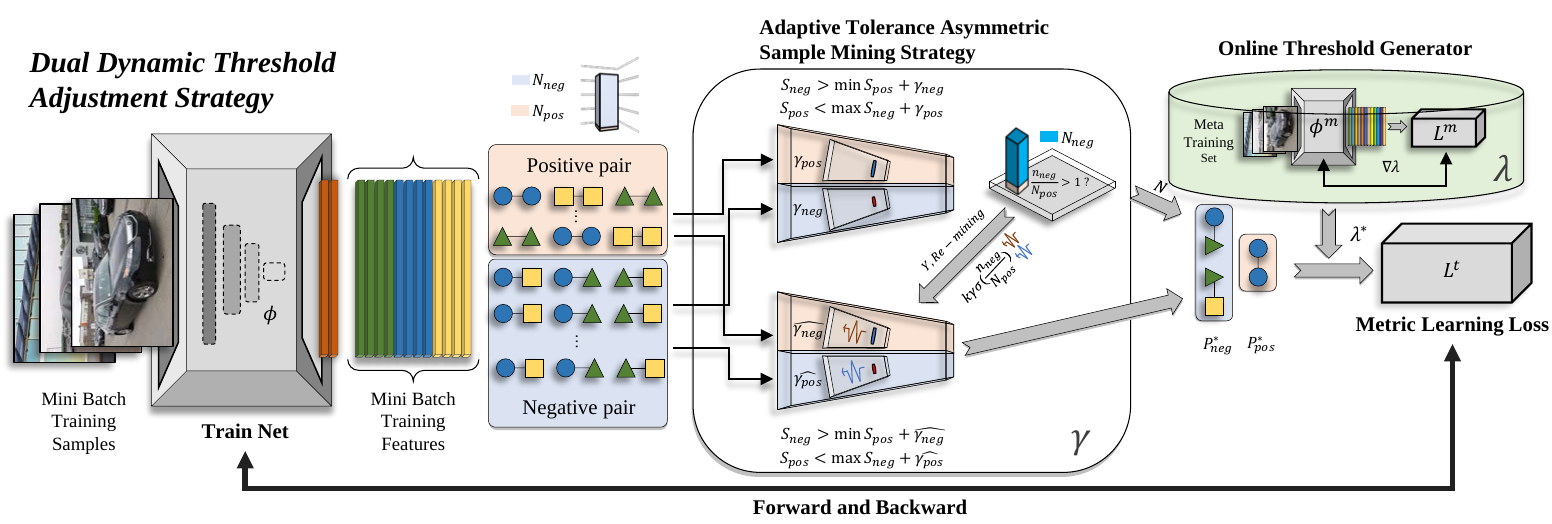}
	\end{center}
	
	\caption{
		\textbf{DDTAS} framework: The sample features output from the backbone is used to form sample pairs.  These sample pairs are all input to the sample mining module.  The sample mining module adopts our proposed Adaptive Tolerance Asymmetric Sample Mining Strategy (AT-ASMS).  This strategy first uses our proposed ASMS (Asymmetric Sample Mining Strategy) to obtain information on the number of sample pairs after initial screening.  If the sample pairs are not uniform or there are too few positive samples, our algorithm will reselect the sample pairs by the adaptively adjusted new threshold $\gamma$.  If the ratio of positive and negative samples is appropriate, these sample pairs directly participate in the following loss function calculation.  Next, before computing the loss function, we use the Online Threshold Generator to obtain a threshold that matches the current training state.  This threshold will serve as the threshold for the loss function.}
	\label{DDTASfrm}
	\vspace{-0.5cm}
\end{figure*}

\textbf{Mining Strategy.} Deep metric learning emphasizes the relationship between samples, commonly utilizing sample pairs or tuples of multiple samples as inputs to its loss functions.
For example, contrastive loss~\cite{Authors16} uses sample pairs as input, while the triplet loss~\cite{Authors21} uses a triplet consisting of an anchor, a positive sample, and a negative sample to form the input of the loss function. In order to obtain richer structural information between samples, various approaches have extended the number of samples within a single input tuple, such as Quadruplet-wise loss~\cite{ni2017fine}, lifted structured feature embedding~\cite{Authors49}, and \textit{N}-pair loss~\cite{Authors59}. Using these loss functions based on sample pairs or tuples must consider the computational cost. If $N$ is the number of samples in the dataset, these algorithms can generate on the order of $O(N^2)$ or $O(N^3)$ and more sample pairs or tuples. 
Therefore, the mining strategy and the weighting algorithm have also become critical steps in deep metric learning.

How to filter out sample pairs with a large amount of information is the key to designing mining strategies. Earlier algorithms obtained more significant gradients by training on the hardest pair of samples~\cite{Authors16,rippel2015metric,Authors49}.
Such strategies enhance training efficiency by selecting highly informative pairs using a fixed margin, effectively utilizing limited computational resources. However, they also possess apparent limitations. Firstly, this overly strict filtering strategy excludes a large number of pairs containing valuable information. Secondly, when there are few samples in the dataset or contain noisy data, this strategy's definition of hard pairs (i.e. $argmin(D_{neg}-D_{pos})$) will cause the training to converge to a minimum local case.
In the work of Schroff \al \cite{Authors21}, the authors used a relatively loose sampling strategy when screening hard pairs to reduce the impact of the most challenging samples. This approach involves conducting hard sample mining under the condition that the distance between negative pairs surpasses the distance between positive pairs and is therefore named semi-hard sampling. 
This strategy sets a fixed threshold. It determines pairs that offer a moderate challenge for the model during training. It also avoids overfitting and slower convergence issues caused by excessively challenging pairs. This helps maintain robustness towards noisy datasets~\cite{sun2022pnp,NPN,10105896,10128961}. However, the strategy's limitation lies in its sensitivity to parameters; inappropriate definition of the range of semi-hard pairs by the threshold can result in poor model performance.
BIER~\cite{Authors58} inherits the features on ~\cite{freund1997decision}, which splits the last larger-dimensional embedding layer of CNNs used in deep metric learning into distinct, non-overlapping smaller learners and iteratively reweights the samples through the gradient of the loss function obtained by training these learners. Harwood \al proposed a more adaptive intelligent mining algorithm~\cite{Authors3,10298026} to increase further the number of valid pairs entering training. This algorithm cooperates with the loss that inherits the triplet and global loss characteristics to obtain better image retrieval performance. PADS~\cite{Authors68} uses the teacher and learner networks to change the sampling distribution dynamically.

\subsection{Meta Learning}
Meta-learning serves two crucial objectives: addressing the need for substantial data in deep learning to achieve proficient prediction performance, and tackling the challenge of limited transferability to new tasks. 
Unlike deep learning, which takes data directly as input, meta-learning takes multiple sub-training tasks as input, each containing sub-datasets. Through the training of subtasks, the meta-learning model acquires prior, subsequently used to guide the training of the main task model, facilitating rapid learning for the main task.
Meta-learning hopes that the main task model can recognize new things better than humans using experience and limited samples.
Since meta-learning is not optimized for specific domains and tasks, the meta-learning framework can be used as long as deep learning algorithms are needed to optimize the main task. Currently, meta-learning is applied to tasks such as zero-shot learning~\cite{verma2020meta}, one-shot learning~\cite{Authors27}, and few-shot learning~\cite{Authors29}. 
Among these tasks, we can train the training task of each dataset containing a small number of training samples as a meta-training task so that the main task can obtain better retrieval effect and generalization performance.
Meta-learning can broadly be categorized into metric-based \cite{Authors29,Authors27,Authors28,mao2023attention}, model-based \cite{Authors30,Authors10,sun2022pnp}, and optimization-based \cite{Authors31,Authors32,Authors12,NPN,yao2020towards}. The work presented in this paper draws inspiration from optimization-based approaches and attempts to optimize the hyperparameter (threshold) in metric learning.

\section{Methodology}\label{sec3}
This section describes each part of our deep metric learning algorithm in detail. In subsection \ref{sec3-4}, we introduce our Dual Dynamic Threshold Adjustment Strategy in detail, and its frame diagram is shown in Fig. \ref{DDTASfrm}.

\subsection{Contrastive Loss}\label{sec3-1}

\begin{figure}[!t]
	\centering{\includegraphics[width=0.98\linewidth]{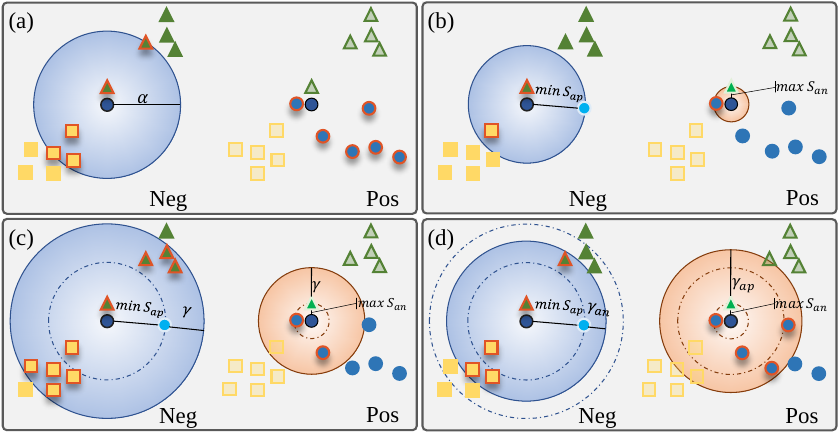}}
	\caption{Sample mining strategy: (a) contrastive loss; (b) semi-fixed sample mining strategy (no tolerance); (c) semi-fixed sample mining strategy (tolerance $\gamma$)  (d) asymmetric sample mining strategy (ours).}
	\label{fig:Similarity distribution map A}
	\vspace{-0.5cm} 
\end{figure}
The contrastive loss~\cite{Authors16} was proposed in Hadsell's work in 2006, significantly enhancing the classic Siamese Network's ability~\cite{Authors14, Authors15} to learn a better embedding space. This loss function aims to minimize the distance between samples of the same class and maximize the distance between samples of different categories.
Specifically, it uses a fixed threshold to separate positive sample pairs and sub-sample pairs to achieve the purpose above, as shown in Fig. \ref{fig:Similarity distribution map A}(a). Let $x_i \in T$ denote the samples in the training set, each paired with a corresponding label $y_i$. The samples pass through the backbone network and are projected as $\Phi(x_i,\theta)$. Then the distance between two samples is computed as: $D:=||\Phi(x_i,\theta)-\Phi(x_j,\theta)||_{l_2}$. $D_{pos}$ represents the distance between samples from the same class, and $D_{neg}$ represents the distance between samples of different classes. The naive form of contrastive loss is as follows: 

\begin{equation}
	\label{E1}
	\begin{aligned}
		\mathcal{L}^{\text {contrastive}}&=\left[\alpha-D_{neg}\right]_{+}+D_{pos}\\ 
	\end{aligned}.
\end{equation}

From the perspective of hyperparameters, besides encouraging the distance between pairs of negative samples to be larger than the threshold $ \alpha $, contrastive loss also hopes that the distance between positive samples approaches 0. In other words, 0 is also a fixed threshold. Since many deep metric learning algorithms use the similarity ($S:=\langle\Phi(x_i,\theta),\Phi(x_j,\theta)\rangle$) between samples instead of distance, we rewrite the formula of contrastive loss as:

\begin{equation}
	\label{E2}
	\begin{aligned}
		\mathcal{L}^{\text {contrastive}}&=\left[S_{neg}-\alpha_{neg} \right]_{+}+\left[\alpha_{pos}-S_{pos}\right]_{+}   \\
	\end{aligned}.
\end{equation}

\subsection{Asymmetric Sample Mining Strategy}\label{sec3-2}
In deep metric learning, the processing of data in combinations of multiple samples is a distinctive feature. This leads to the generation of a substantial number of sample tuples(e.g., pairs $O(N^2)$, triples $O(N^3)$). Putting so many sample units into training cannot quickly converge, and it is also a waste of computing resources. Extensive research has demonstrated the significance of employing effective sample mining strategies to select informative samples or sample pairs from large datasets. They can eliminate a large number of sample pairs with low information content to accelerate convergence, improving retrieval performance and reducing computational consumption.

The LMNN~\cite{Authors48} and Triplet Loss~\cite{Authors21} inspire us. Specifically, we use the relative similarity between positive and negative pairs to extract crucial samples. For the same anchor, we treat the negative samples with greater similarity than the most difficult positive samples (farthest from the anchor) as informative negatives. Similarly, positive samples with the potential to be mined need to be farther than negative samples closest to the anchor. At the same time, the relative similarity brings the variable size capability to the mining strategy. The formulated condition for mining positive pairs is as follows:
\begin{equation}
	\label{E4}
	\begin{aligned}
		S_{pos}&< \max S_{neg}. 
	\end{aligned}
\end{equation}
For negative pairs, their filtering condition is:
\begin{equation}
	\label{E5}
	\begin{aligned}
		S_{neg}&> \min S_{pos}. 
	\end{aligned}
\end{equation}

According to Eq. (\ref{E2}), we know that the sample pair mining conditions for the contrastive loss are $ S_{pos} < \lambda_{pos} $ (previous studies tend to set $\lambda_{pos}$ to 0) and $ S_{neg} > \lambda_{neg} $ respectively, and their form is similar to Eq. (\ref{E4}) and Eq. (\ref{E5}). As shown in Fig. \ref{fig:Similarity distribution map A}(b), this sample mining strategy uses unfixed maximum anchor-negative similarity and a variable minimum anchor-positive similarity to filter informative samples, which increases the adaptability of the algorithm to different data. However, using the similarity of the hardest sample pair as a threshold will cause too few sample pairs to meet the mining conditions in the later stage of training. 
In contrast, the MS loss~\cite{Authors7} employs a similar strategy but increases the tolerance of screening conditions by threshold $\gamma$, as shown in Fig. \ref{fig:Similarity distribution map A}(c). This method increases the number of sample pairs integrated into the training process:
\begin{equation}
	\label{E6}
	\begin{aligned}
		S_{pos}&< \max S_{neg} + \gamma.  
	\end{aligned}
\end{equation}
For anchor-negative pairs, their filtering condition is:
\begin{equation}
	\label{E7}
	\begin{aligned}
		S_{neg}&> \min S_{pos} - \gamma.
	\end{aligned}
\end{equation}

If the mining strategy uses the same threshold for both positive and negative sample pairs, the number of positive pairs tends to be very small. This exacerbates the imbalance of sample pairs. Our proposed Asymmetric Sample Mining Strategy (ASMS) uses two different threshold to solve this problem:

\begin{equation}
	\label{E8}
	\begin{aligned}
		S_{pos}&< \max S_{neg} + \gamma_{pos}.   
	\end{aligned}
\end{equation}
\begin{equation}
	\label{E9}
	\begin{aligned}
		S_{neg}&> \min S_{pos} - \gamma_{neg}.   
	\end{aligned}
\end{equation}

It can be seen from the formula and Fig. \ref{fig:Similarity distribution map A}(d) that this strategy can dynamically adjust the amount of mined samples by using the hardest sample pairs in different training stages. However, as the proximity between samples within a class diminishes, employing a larger fixed threshold could admit numerous uncomplicated samples into training, potentially impeding the gradient descent process due to redundant, simplistic sample information. Conversely, if the threshold is too small, there will be too few valid sample pairs, and the training will fall into a local optimum. Therefore, to address this issue, we further optimize our ASMS. Optimization details will be described in subsequent subsections.

\subsection{Soft Contrastive Loss}\label{sec3-3}
The mined informative samples are used to calculate the loss. In deep metric learning, existing algorithms often fail to distinguish the importance of pairs after mining, particularly those near the decision boundary, posing a significant limitation. To tackle this challenge, we introduce a novel Soft Contrastive loss function that leverages the binomial distribution's characteristics to finely differentiate the importance of mined pairs, a capability lacking in traditional loss functions. We take Eq. (\ref{E2}) as the basic form of the loss function. At the same time, inspired by the LSE algorithm and the binomial deviance loss, we use the softplus function to replace the hinge function of the contrastive loss. Its form is as follows:

\begin{equation}
	\label{E10}
	\begin{aligned}
		\mathcal{L}_{scon}&= \frac{1}{N_t}\sum\left\{ \frac{1}{\mu N_{pos}}\sum_{pos}\log\left[1 + e^{\mu\left(\lambda-S_{pos}\right)}\right]\right. \\
		&\left. +\frac{1}{\nu N_{neg} }\sum_{pos}\log\left[1 + e^{\nu\left(S_{neg}-\lambda\right)}\right] \right\}
	\end{aligned}.
\end{equation}

$N_t$ represents the amount of data in the training set and $\lambda$, $\mu$, $\nu$ are hyperparameters, where the threshold $\lambda$ is the focus of our work. $\lambda$ uses absolute distance to separate positive and negative pairs, thereby influencing the distribution of sample similarity.  Similar to the threshold in the mining strategy, due to the change of the embedding space, the threshold also requires adaptive adjustment during training.  Therefore, we use the meta-learning strategy to turn the hyperparameters $\gamma_{pos}$, $\gamma_{neg}$ and $\lambda$ into adaptable parameters learned dynamically.  We will elaborate this in the next subsection. 

Our Soft Contrastive loss function assigns distinctive weights to each pair, allowing the algorithm to concentrate more on critical sample pairs during model training. This innovative approach significantly enhances the discriminative power of learned features, improving model accuracy and offering new insights into deep metric learning.

\subsection{Dual Dynamic Threshold Adjustment Strategy}\label{sec3-4}
It is inefficient to employ manual tuning for hyperparameters, so we further convert these hyperparameters into automatically tunable parameters. For the two sample mining thresholds, $\gamma_{pos}$ and $\gamma_{neg}$, which cannot be integrated into the deep learning computation graph, we design a new threshold adjustment strategy aimed at mitigating the severe imbalance between positive and negative pairs. In the context of the loss function, we use the meta-method to enable the network to dynamically adjust $\lambda$ according to the current training state.

\begin{figure}[t]
	\begin{center}
		\includegraphics[width=1\linewidth]{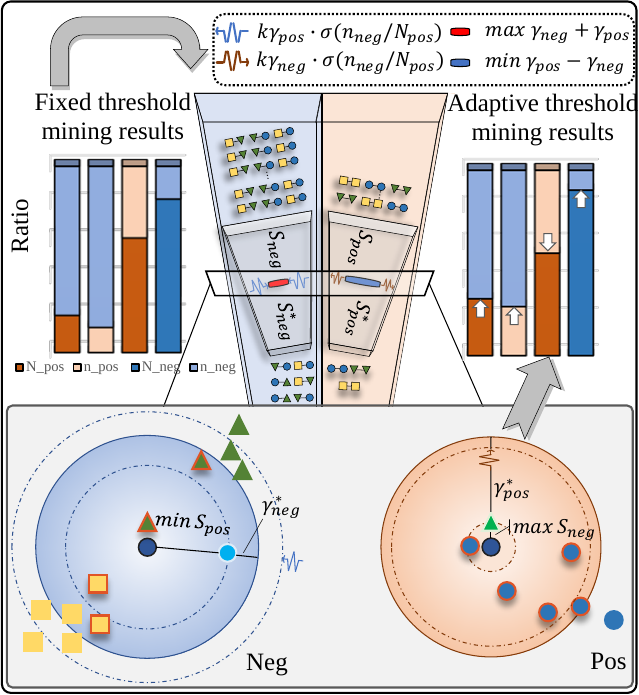}
	\end{center}
	\vspace{-0.3cm}
	\caption{Adaptive Tolerance Asymmetric Sample Mining Strategy }
	\label{frmmine}
	\vspace{-0.5cm} 
\end{figure}
\subsubsection{Adaptive Tolerance Asymmetric Sample Mining Strategy}
The incorporation of the hardest positive and negative pairs in Eq. (\ref{E4}) to Eq. (\ref{E9}) make the mining strategies for screening informative sample pairs highly adaptable. The parameters $\gamma_{pos}$ and $\gamma_{neg}$ are purposefully designed to offer distinct, fixed tolerances for screening positive and negative pairs. This tolerance can either relax the mining conditions to allow subsequent steps to obtain more sample pairs or, alternatively, tighten the sample screening area ($\gamma_{pos}<0$ or $\gamma_{neg}<0$) for getting more challenging instances. In practical cases, the number of negative pairs tends to be substantial. Even if a stricter threshold is used for the negative pair than the positive pair screening conditions, in the early and middle stages of training, many negative pairs still meet the screening conditions. Specifically, within each minibatch, we will load $N_{instance}$ samples of each class for $\frac{B}{NI}$ (B-Batchsize) classes. The total number of negative pairs is:

\begin{equation}
	\label{E11}
	\begin{aligned}
		N_{neg} = \frac{1}{2}\left(B^2-B\cdot N_{instance}\right).
	\end{aligned}
\end{equation}

For positive sample pairs, their number is naturally scarce. Its total amount is:

\begin{equation}
	\label{E12}
	\begin{aligned}
		N_{pos} = \frac{1}{2}\left(B\cdot N_{instance}-B\right).
	\end{aligned}
\end{equation}

For example, when the batch size is 80, we calculate that the number of positive pairs is 160 and negative pairs are 3,000. In deep metric learning algorithms, particularly with a small batch size, mitigating the dearth of training information caused by insufficient instances in the later stages of training proves challenging even with weighting operations. We hope that $\lambda$ can expand the mining range of positive pairs according to the quantitative relationship between positive and negative pairs in the current training state and, at the same time, screen negative pairs more finely. Therefore, we design a floating tolerance term to cooperate with $\gamma_{pos}$ and $\gamma_{neg}$ to achieve the aforementioned purpose. The two dynamic $\gamma$ are formulated as:

\begin{equation}
	\label{E13}
	\begin{aligned}
		\hat{\gamma_{pos}} = \gamma_{pos} +\kappa\gamma_{pos}{Sigmoid}\left(\frac{n_{neg}}{N_{pos}}\right),
	\end{aligned}
\end{equation}

\begin{equation}
	\label{E14}
	\begin{aligned}
		\hat{\gamma_{neg}} = \gamma_{neg} -\kappa\gamma_{neg}{Sigmoid}\left(\frac{n_{neg}}{N_{pos}}\right),
	\end{aligned}
\end{equation}
where $\kappa$ represents a hyperparameter. The algorithm flow of the adaptive tolerance sample mining strategy is as shown in Alg. \ref{alg:atsm}.

\begin{algorithm}[t]
	\caption{$Adaptive\ Tolerance\ Asymmetric\ \\ Sample\  Mining\ Strategy$} 
	\label{alg:atsm} 
	\begin{algorithmic}[1] 
		\REQUIRE ~~\\ 
		Training sample pairs (mini-batch): $P_{pos},P_{neg}$;\\
		Mining Strategy: $M(P_{\text{train}},\gamma_{a},\gamma_{b})$; \\
		Total number of positive pairs: $N_{pos}$; \\
		The zoom parameter: $\kappa$\\
		Fixed threshold: $\gamma_{pos},\gamma_{neg}$\\
		Sigmoid function: $\sigma(\xi)$
		\ENSURE ~~\\ 
		The new threshold: $\hat{\gamma_{pos}},\hat{\gamma_{neg}}$;\\
		The new training sample pairs(mini-batch): ${p_{pos}}^*,{p_{neg}}^*$;\\
		\STATE $n_{pos},n_{neg} = M(P_{pos},P_{neg},\gamma_{pos},\gamma_{neg})$\\
		\STATE $\xi=\frac{n_{neg}}{N_{pos}}$
		\IF{$\xi>1$} 
		\STATE $\epsilon_{pos} = \kappa\gamma_{pos}\cdot\sigma(\xi);\epsilon_{neg} = \kappa\gamma_{neg}\cdot\sigma(\xi) $\\
		\STATE $\hat{\gamma_{pos}} = \gamma_{pos} +\epsilon_{pos};\hat{\gamma_{neg}} = \gamma_{neg} -\epsilon_{neg}$\\
		\ELSE
		\STATE $\hat{\gamma_{pos}} = \gamma_{pos};\hat{\gamma_{neg}} = \gamma_{neg}$
		\ENDIF 
		\STATE $p_{pos}^*,p_{neg}^* = M(P_{pos},P_{neg},\hat{\gamma_{pos}},\hat{\gamma_{neg}})$\\
	\end{algorithmic}
\end{algorithm}

As shown in Fig. \ref{frmmine}, in each iteration, we first perform a sample mining strategy with fixed thresholds $\gamma_{pos}$ and $\gamma_{neg}$ to get the number of positive and negative pairs we need. Then we calculate the ratio of mined negative samples ($n_{neg}$) to the total positive samples, denoted by $\xi$. Next, we use $\sigma$ to assess whether there exists a significant disparity in quantity between positive and negative pairs. When the sample pair meets this assessment condition (see step 3 in Alg. \ref{alg:atsm}), it indicates that the positive pair is in a state of being rare. At the same time, this state is also accompanied by an imbalance between positive and negative pairs. Therefore, if the result is `yes' (see step 3 in Alg. \ref{alg:atsm}), a new threshold is calculated using Eq. (\ref{E13}) and Eq. (\ref{E14}), thereby relaxing the positive pair screening conditions and tightening the negative sample pair screening conditions (see step 3-5 in Alg. \ref{alg:atsm}). Through this operation, the number of positive sample pairs participating in training can be increased, and the uneven proportion of the two kinds of sample pairs can be alleviated. If the results do not meet the judgment conditions (see step 3 in Alg. \ref{alg:atsm}), the training has entered a later stage, and most samples have already reached a reasonably balanced embedding space. At this time, the pairs that have passed step 1 in Alg. \ref{alg:atsm}) contain many informative sample pairs. Therefore, we no longer augment or reduce these sample pairs in this case.
Section \ref{sec3-2} compare the mining strategies employed by our algorithm and MS Loss~\cite{Authors7}, both utilizing thresholds as tolerances to filter samples. In contrast to MS Loss, which employs a single fixed threshold to filter positive and negative pairs separately, our algorithm offers two improvements: Firstly, we utilize two distinct thresholds for filtering based on the disparity in quantity between positive and negative pairs. Secondly, we employ dynamically changing thresholds during training for adaptive mining.

\subsubsection{Loss function threshold adjustment algorithm}
After the mining step, sample pairs endowed with richer information are incorporated into the loss function. To enable our proposed sample mining strategy with the ability to adjust parameters automatically, we propose a meta-learning-based loss function threshold adjustment algorithm and introduce a corresponding meta-learning threshold generator.
Our algorithm inherits some properties of the algorithm~\cite{Authors13} proposed by Ren Mengye to reweight samples using meta-learning. This work adds a meta-learning module to the normal flow of deep learning algorithms that require sample weighting. The distinctive aspect of our approach is that it focuses on pairs of samples rather than individual samples. Moreover, our task is to assign thresholds to sample pairs based on training stages, as opposed to assigning weights as in~\cite{Authors13}. Next, we introduce the framework and specific process of the Meta-learning-based loss function threshold adjustment algorithm.\\

\textbf{Meta Training Set.}
Inspired by~\cite{Authors13}, we recognize the importance of a validation set consisting of clean and unbiased data as the input to the meta-learning module. This validation set serves to alleviate the problem of noisy training samples. The training set does not suffer from data inhomogeneity and noise in our retrieval task. Our goal is to leverage a lightweight data set to enable dynamic adjustment of the mining strategy in the regular training module and the threshold in the loss function. Therefore, we only require that this small data set contains samples of all categories in the training set. We refer to this dataset as the meta training set. To further improve the ease of implementation of the algorithm, we use the data in the training set as the source for the meta training set. Note that the meta training set data is included in the training set.

\textbf{Online Threshold Generator.} We take the threshold in the loss function as a learnable parameter and propose an online threshold generator that inherits meta-learning features to provide adaptive thresholds for sample pair-based loss functions. The backbone used in the meta-learning process is consistent with that used in the main training process. When the threshold generator is in operation, we load the backbone network parameters of the current regular training module into the meta-learning threshold generator backbone. Subsequently, we leverage we use the gradients generated by training the meta-learning set to derive dynamic non-artificial threshold. One of the most straightforward ways is to nest the two optimization loops (i.e., regular training and meta-learning) to obtain the optimal threshold. To facilitate the representation of the optimization process, we first simplify the input of the loss function, that is, the sample pairs that have passed the mining strategy (step 9 at Alg. \ref{alg:atsm}):\\
\begin{equation} 
	\label{eq::mine_strategy_brf}
	p_{pn} = M(P_{pn},\hat{\gamma_{pn}}).
\end{equation}
At this time, the nested optimization is expressed as:
\begin{equation} 
	\label{eq::meta_mg}
	\lambda^*=\arg \min _{\lambda}\sum \mathcal{L}^{\text{m}}\left(p_{pn}^{mts}, \arg \min _{\theta} \mathcal{L}^{\text{t}}(\lambda, p_{pn} | \theta)\right),
\end{equation}
where $\mathcal{L}_{t}$ and $\mathcal{L}_{m}$ represent the loss function in the main training and online threshold generator, respectively.\\

\begin{table*}[t]
	\centering
	\renewcommand{\arraystretch}{1.5}
	\caption{Recall@$K$(\%) and NMI performance on CUB200 and Cars196. Abbreviations for threshold Characteristics: Fix-Fixed, Semifix-Semifixed, Sym-Symmetry, Asym-Asymmetry, Dyn-Dynamic. }
	\label{AS1}
	\resizebox{\textwidth}{!}{
		\begin{tabular}{l|cc|c|cccccc|c|cccccc}
			\hline
			\multirow{2}{*}{\textbf{Recall@$K$(\%)\qquad\qquad}}& \multicolumn{2}{c|}{\textbf{Threshold Characteristic}} & \multicolumn{7}{c|}{\textbf{CUB200}} & \multicolumn{7}{c}{\textbf{Cars-196}} \\
			\ & Mining & Loss & NMI & 1 & 2 & 4 & 8 & 16 & 32 & NMI &1 & 2 & 4 & 8 & 16 & 32 \\
			\hline
			\multirow{1}{*}{\textbf{MS+Sim}$_p$~\cite{Authors7}} & Semifix/Sym & Fix/Sym
			& - & 65.7 &  77.0 & \textbf{86.3} & 91.2 & 95.0 & 97.3 & - & 84.1 & 90.4 & 94.0 & 96.5 & 98.0 & 98.9 \\
			\multirow{1}{*}{\textbf{Soft Con+Sim}$_p$} & Semifix/Sym & Fix/Sym
			& 69.3 &  65.9 & 77.2 & 85.9 & 91.1 & 95.2 & 97.6 & 71.0 & 84.4 & 91.0  &  94.3  & 96.4  & 97.9 & 98.9\\
			\multirow{1}{*}{\textbf{MS+Sim}$_p$ ($\gamma={0}$)} & Semifix/Asym & Fix/Sym 	
			& 69.9 & 66.0 &  77.5 & 85.9 & 92.0 & 95.1 & 96.2 & 71.0 & 84.7 & 91.0 & 94.2 & 96.9 & 98.1 & 99.0 \\	
			\multirow{1}{*}{\textbf{Soft Con+Sim}$_p$ ($\gamma={0}$)} & Semifix/Asym & Fix/Sym 	
			& 70.3 & 66.3 &  78.0 & 86.4 & 92.4 & 95.9 & 97.0 & 71.5 & 85.1 & 91.3 & 94.4 & 96.0 & 98.3 & 99.1 \\	
			\multirow{1}{*}{\textbf{MS + ASMS (ours)}} 	& Semifix/Asym & Fix/Sym
			& 70.5 &  67.4 & 78.2 & 84.7 & 89.6 & 95.1 & 96.0 & 71.8 & 84.7 & 90.8  & 94.5  & 96.7  & 97.9 & 99.0 \\ 
			\multirow{1}{*}{\textbf{Soft Con + ASMS (ours)}} 	& Semifix/Asym & Fix/Sym
			& 70.8 &  68.0 & 76.9 & 85.9 & 91.2 & 95.6 & 97.3 & 73.4 & 85.4 & 91.2  & 94.7  &  97.2  &  98.5 &  99.0\\ 
			\multirow{1}{*}{\textbf{MS + AT-ASMS (ours)}} 	& Dyn/Asym & Fix/Sym
			& 70.4 & 67.5 & 78.1 & 85.5 & 90.9 & 95.0 & 96.7 &  72.7 &  85.4 & 91.3  & 94.8  & 96.8  & 97.5 & 99.2\\
			\multirow{1}{*}{\textbf{Soft Con + AT-ASMS (ours)}} 	& Dyn/Asym & Fix/Sym
			& \textbf{71.1} & \textbf{68.3} & \textbf{78.8} & \textbf{86.2} & \textbf{91.7} & \textbf{95.6} & \textbf{97.9} & \textbf{73.3} & \textbf{86.4} & \textbf{92.0}  &\textbf{95.4}  &\textbf{97.2}  & \textbf{98.5} & \textbf{99.2}\\
			\hline
			
			\multirow{1}{*}{\textbf{Soft Con$^*$+ Sim$_p$} ($\gamma={0}$)} & Semifix/Asym & Dyn/Sym 	
			& 70.0 & 66.3 &  77.8 & 86.3 & 92.2 & 95.9 & 97.3 & 71.7 & 85.3 & 91.2 & 94.5 & 96.6 & 98.4 & 99.0 \\	
			\multirow{1}{*}{\textbf{Soft Con$^*$+Sim$_p$}} & Semifix/Sym & Dyn/Sym
			& 69.2 & 65.9 &  77.4 & 86.3 & 91.7 & 94.7 & 97.0 & 71.3 & 84.8 & 91.0 & 94.6 & 96.8 & 98.3 & 98.9 \\	
			\multirow{1}{*}{\textbf{Soft Con$^*$ + ASMS }} 	& Semifix/Asym & Dyn/Sym
			& 70.4 & 68.2 & 78.1 & 86.5 & 92.9 & 95.1 & 97.4 & 73.0 & 85.9 & 91.7  & 95.0  & 97.2  & 98.3 & 99.2 \\ 
			\hline
			\multirow{1}{*}{\textbf{DDTAS (i.e., Soft Con$^*$ + AT-ASMS}) } 	& Dyn/Asym & Dyn/Sym
			& \textbf{71.0} & \textbf{68.4} & \textbf{78.7} & \textbf{86.7} & \textbf{92.1} & \textbf{95.6} & \textbf{97.7} & \textbf{73.3} & \textbf{86.4} & \textbf{92.0}  &\textbf{95.4}  &\textbf{97.2}  & \textbf{98.5} & \textbf{99.2}\\
			\hline
	\end{tabular}}
	
\end{table*}

The inner loop of Eq. (\ref{eq::meta_mg}) represents the regular training process aimed at determining the optimal network parameters. Once this inner loop successfully identifies find the optimal network parameters, they are stabilized and transferred to serve as the backbone for the meta-learning module network. Then we use the meta training set to perform a complete iterative optimization to find the optimal threshold. But completing the alternating operation of such large loops demands considerable time and computational resources. To address this issue, we configure the online threshold generator as a one-step look ahead mode, wherein it conducts a single gradient descent operation on a mini-batch. By doing this, we can get the estimated threshold quickly. Specifically, we use stochastic gradient descent (SGD) in each iteration of each regular training to get the network parameters for the next step:

\begin{equation}
	\label{eq::theta_lmbda}
	\hat{\theta}_{t+1}(\hat{\lambda})=\theta_{t}-\psi \cdot \sum \frac{\partial \mathcal{L}^t\left.(\hat{\lambda}, p_{pn},\theta)\right|_{\theta=\theta_{t}}}{\partial\theta},
\end{equation}
where $\psi$ is the learning rate. We estimate $\lambda$ with our online threshold generator at step $t$:
\begin{equation}
	\begin{aligned}
		\hat{\lambda}_t \approx  \left[- \varphi \cdot \left.\frac{\partial}{\partial \lambda_t}\mathcal{L}^m\left(p_{pn}^{mts},\arg \min_{\theta} \sum \mathcal{L}^t(\lambda, p_{pn} | \theta)\right)\right|_{\lambda =\lambda_t}\right]_+
		\label{eq:approx}
	\end{aligned},
\end{equation}
where $\varphi$ is the descent step size. And $\hat{\lambda}_t$ is the threshold we get through the guidance of each sample type in the meta training set.


\section{Experiments}\label{sec5}
\subsection{Datasets and Experimental Settings}

\textbf{CUB200-2011 (CUB200)} contains 11,788 images of 200 bird species. In the experiment, we split it into two parts. All 5,864 images in the first 100 categories of the dataset are applied for model training. The 5,924 images in the remaining 100 categories are devoted to demonstrating the retrieval performance of the model.\\
\textbf{Cars196} has a total of 16,185 pictures of different models of cars, categorized into 196 classes. We utilize the images of the first 98 models of cars as the training set and the last 98 models as the test set.\\
\textbf{Stanford Online Products (SOP).} Unlike the aforementioned two datasets, Stanford Online Products is a large-scale dataset for few-shot image retrieval with few samples per class. The dataset contains 120,053 online product images in 22,634 categories. It contains 59,551 images of 11,318 categories, while 60,502 images from the remaining 11,316 categories are for testing. \\
\textbf{Experimental Settings.} In this subsection, we outline the standard settings utilized in the subsequent experiments. Referring to \cite{Authors7}, we first resize all input images to 256$ \times $256 and crop them to 224$ \times $224. Meanwhile, according to \cite{Authors7}, we apply image augmentation on these pictures during the training phase. We only use the center crop operation in the testing phase to process the image. When loading data in each minibatch, we randomly select $N=5$ instances from $\frac{Batchsize}{N}$ categories of data. Preprocessed training images are fed into the backbone. We use BN-Inception as our backbone network, which is pretrained on ImageNet. And we use Adam as the optimizer. In the experiments, the learning rate is set to $10^{-5}$, and the embedding dimension is 512. Sample features output from the backbone are processed with $L_2$ normalized. The hyperparameters in the experiments are set as follows: $\lambda=0.7$, $\mu=2$, $\nu=40$, $\kappa=0.5$, $\gamma_{pos}=0.1$, and $\gamma_{neg}=0.01$. All our experiments are conducted on a 24G NVIDIA TITAN RTX GPU. To report the performance of our proposed method, we follow previous work and choose Recall@K (\%) and Normalized Mutual Information (NMI) as metrics to demonstrate the algorithm's performance.

\begin{figure*}[t]
	\centering
	\begin{center}
		\includegraphics[width=1\linewidth]{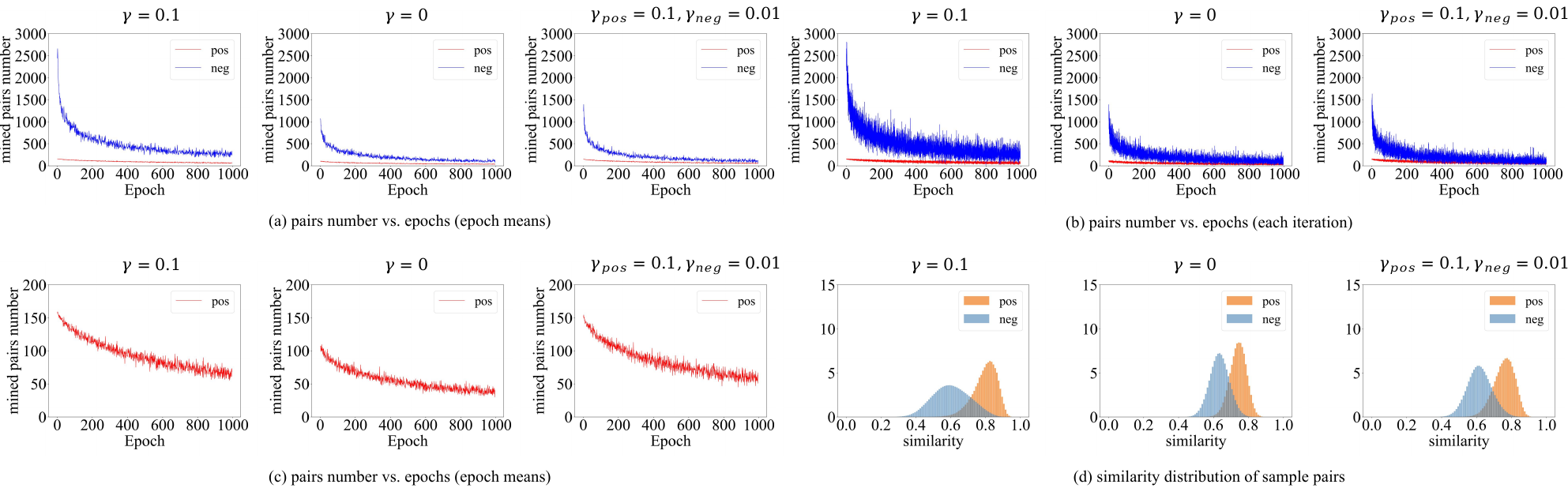}
	\end{center}
	\caption{The figure of the relationship between positive and negative pairs in the training process. Subfigures (a), (b), and (c) describe the change in the number of positive and negative pairs. The red line represents the positive pair, and the blue line represents the negative pair. Subfigure (a) is the mean of the number of pairs for multiple optimization iterations in each epoch. It is used to analyze overall volume trends. Subfigure (b) represents the specific number of pairs in each iteration, which is used to observe the actual state of the two types of pairs. Subfigure (c) shows the change in the number of positive pairs during training. Subfigure (d) is the similarity distribution map of the pair corresponding to the best retrieval result. Orange and blue correspond to positive and negative pairs, respectively.}
	\label{pncmp}
\end{figure*}

\subsection{Ablation Studies}
\label{as}
\subsubsection{Image Retrieval Performance Comparison}\label{as1} We conduct ablation studies to analyze the impact of various parts of our algorithm on the experimental results of image retrieval. We use MS Loss~\cite{Authors7} as the reference loss for ablation studies.\\
\textbf{Mining Strategy}: As can be seen from Table~\ref{AS1}, when we use the sample mining strategy based on Similarity-P \cite{Authors7} for the loss function, the image retrieval performance of the algorithm is unsatisfactory due to the use of a symmetric, large threshold. When we set the threshold of Similarity-P to 0, the algorithm demonstrated superior retrieval performance compared to employing a larger symmetric threshold. This experiment underscores the substantial improvement in classification and clustering performance achievable by imposing stringent constraints on negative samples. However, it also reveals the problem of symmetric threshold still exists. Too strong constraints lead to too few positive samples that can enter the training, which significantly reduces the information that the model can learn, thus hindering the further improvement of the algorithm. Thus, we used our proposed Asymmetric Sample Mining Strategy (ASMS) to replace the Similarity-P mining strategy. The ASMS provides different thresholds $\gamma_{pos}$ and $\gamma_{neg}$ for the positive and negative pairs to be screened ($\gamma_{pos}>\gamma_{neg}$). As can be seen from Table~\ref{AS1}, ASMS yields superior performance. Compared with the algorithm using Similarity-P, the Recall@1 of our ASMS is improved by 2.4\% ($65.9\% \rightarrow 68.3\%$) on CUB200 and improved by 2.0\% ($84.4\% \rightarrow 86.4\%$) on Cars196. This result shows that providing greater tolerance for a smaller number of positive samples and implementing stronger constraints for negative samples can improve the algorithm's image retrieval performance. Subsequently, we employed the Adaptive Tolerance Asymmetric Sample Mining Strategy (AT-ASMS) which enables our ASMS to adjust thresholds dynamically. We can see that the retrieval algorithm effect has also been substantially improved.  Compared with the Similarity-P mining strategy, AT-ASMS improves Recall@1 (\%) by 2.5\% and NMI by 1.7\% on CUB200.\\
\textbf{Loss Function:} The above experiments are done with a fixed threshold in the loss function, corresponding to the upper part of Table~\ref{AS1}. Upon modifying the threshold in the loss function with values generated from the online threshold generator, the algorithm continued to perform admirably across all datasets.

\subsubsection{Visualize Experimental Data}
In addition to presenting numerical results, we also use figures to analyze why the algorithm improves image retrieval performance. Each subfigure in Fig. \ref{pncmp} shows the variation of sample pairs throughout the training under the three sample mining threshold states, respectively. The threshold strategy, originating from MS Loss~\cite{Authors7}, corresponding to the first column in each subfigure provides equal thresholds for positive and negative pairs, that is, to filter sample pairs with the same tolerance. This strategy initially introduces a relatively large number of trainable pairs to the algorithm. However, in the later stage of training, it has the problem of insufficient constraint on negative pairs. The mining strategy shown in the second column is a particular state derived from the first column. This strategy only uses the highest similarity among positive/negative pairs to mine relatively hard negative/positive pairs (rows 3 and 4 in Table~\ref{AS1}). This threshold mining strategy excels in uncovering more informative pairs compared to hard mining. However, this strategy is still somewhat stringent for the screening conditions of sample pairs. In the deep learning training process, it will still indiscriminately remove more valid sample pairs, impeding the model's capacity to acquire crucial information. This third column is our ASMS. This algorithm employs distinct filtering thresholds for positive and negative sample pairs. By assigning a higher threshold for positive pairs and a smaller threshold for negative pairs, the long-term unevenness of positive and negative pairs in training and the insufficient number of positive pairs is alleviated.

\begin{table*}[t]
	\centering
	\renewcommand{\arraystretch}{1}
	\caption{Recall@$K$ (\%) and NMI performance on CUB200 and Cars-196. Each abbreviation corresponds as: BB: Backbone Network, G: GoogleNet, BN-I: BN–Inception, R-50: ResNet50.}
	\label{CMP1}
	\resizebox{1\textwidth}{!}{
		\begin{tabular}{l|c|c|ccccccc|cccccccc}
			\hline
			\multicolumn{1}{c|}{\multirow{2}{*}{\textbf{Method}}}& \multicolumn{1}{c|}{\multirow{2}{*}{\textbf{Dim}}} & \multicolumn{1}{c|}{\multirow{2}{*}{\textbf{BB}}} & \multicolumn{7}{c|}{\textbf{CUB200}} & \multicolumn{7}{c}{\textbf{Cars-196}} \\
			\ & & & NMI & 1 & 2 & 4 & 8 & 16 & 32 & NMI & 1 & 2 & 4 & 8 & 16 & 32\\		
			\hline
			\multirow{1}{*}{LiftedStruct~\cite{Authors49}} 		
			& - & G  & 56.5 & 47.2 & 58.9 & 70.2 & 80.2 & 89.3 & 93.2 & 56.9 & 49.0 & 60.3 & 72.1 & 81.5 & 89.2 & 92.8 \\	
			\multirow{1}{*}{\textit{N}-Pair-Loss~\cite{Authors56}} 		
			& 64 &G & 57.2 & 51.0 & 63.3 & 74.3 & 83.2 & - & - & 57.8 & 71.1 & 79.7 & 86.5 & 91.6 & - & - \\
			\multirow{1}{*}{LoOp~\cite{vasudeva2021loop}} 		
			& 512 &G & 61.1 & 52.0 & 64.3 & 75.0 & 84.1 & - & - & 63.0 & 72.6 & 81.5 & 88.4 & 92.8 & - & - \\
			\multirow{1}{*}{HDC~\cite{Authors57}} 		
			& 384 & G& - & 53.6 & 65.7 & 77.0 & 85.6 & 91.5 & 95.5 & - & 73.7 & 83.2 & 89.5 & 93.8 & 96.7 & 98.4 \\
			\multirow{1}{*}{BIER~\cite{Authors58}} 		
			& 512 & G  & - & 55.3 & 67.2 & 76.9 & 85.1 & 91.7 & 95.5 & - & 78.0 & 85.8 & 91.1 & 95.1 & 97.3 & 98.7 \\
			\multirow{1}{*}{A-BIER~\cite{Authors22}} 		
			& 512 & G  & - & 57.5 & 68.7 & 78.3 & 86.2 & 91.9 & 95.5 & - & 82.0 & 89.0 & 93.2 & 96.1 & 97.8 & 98.7 \\
			
			\multirow{1}{*}{ABE~\cite{Authors59}} 		
			& 512 & G & - & 60.6 & 71.5 & 79.8 & 87.4 & - & - & - & 85.2 & 90.5 & 94.0 & 96.1 & - & - \\
			
			\multirow{1}{*}{XBM~\cite{Authors50}} 		
			& 512 & G & - & 61.9 & 72.9 & 81.2 & 88.6 & 93.5 & 96.5 & - & 80.3 & 87.1 & 91.9 & 95.1 & 97.3 & 98.2 \\
			\hline      
			\multirow{1}{*}{Margin~\cite{Authors44}} 		
			& 128 & R50 & 69.0 & 63.6 & 74.4 & 83.1 & 90.0 & 94.2 & - & 69.1 & 79.6 & 86.5 & 91.9 & 95.1 & 97.3 & - \\
			\multirow{1}{*}{MIC~\cite{Authors68}} 		
			& 512 & R50 & 69.7 & 66.1 & 76.8 & 85.6 & - & - & - & 68.4 & 82.6 & 89.1 & 93.2 & - & - & - \\
			\multirow{1}{*}{PADS~\cite{Authors68}} 		
			& 512 & R50 & 69.9 & 67.3 & 78.0 & 85.9 & - & - & - & 68.8 & 83.5 & 89.7 & 93.8 & - & - & - \\
			\hline
			\multirow{1}{*}{Clustering~\cite{Authors54}} 		
			& 64 & BN-I & 59.2 & 48.2 & 61.4 & 71.8 & 81.9 & - & - & 59.0 & 58.1 & 70.6 & 80.3 & 87.8 & - & -  \\
			\multirow{1}{*}{ProxyNCA ~\cite{Authors55}} 		
			& 64 & BN-I & 59.5 & 49.2 & 61.9 & 67.9 & 72.4 & - & - & 64.9 & 73.2 & 82.4 & 86.4 & 87.8 & - & - \\
			\multirow{1}{*}{\textbf{DDTAS}} 		
			& 64 & BN-I  & \textbf{67.3} & \textbf{59.6} & \textbf{71.3} & \textbf{80.4} & \textbf{88.3} & \textbf{93.5} & \textbf{96.5} & \textbf{67.8} & \textbf{77.4} & \textbf{85.4}  &\textbf{90.8}  &\textbf{94.4}  & \textbf{96.8} & \textbf{98.5}\\
			\multirow{1}{*}{ALA~\cite{Authors62}} 		
			& 512& BN-I & 66.3 & 61.6 & 73.9 & 83.1 & 89.7 & - & - & 68.5 & 80.5 & 87.9 & 92.8 & 95.9 & - & - \\
			\multirow{1}{*}{BD~\cite{liang2021dynamic}}  		
			& 512 & BN-I & - & 61.8 & 73.3 & 83.0 & 89.6 & 94.0 & 96.9 & - & 75.7 & 84.4 & 90.6 & 94.8 & 97.2 & 98.8 \\
			\multirow{1}{*}{SoftTriple~\cite{Authors60}} 		
			& 512 & BN-I & - & 65.4 & 76.4 & 84.5 & 90.4 & - & - & - & 84.5 & 90.7 & 94.5 & 96.9 & - & - \\
			\multirow{1}{*}{MS~\cite{Authors7}} 		
			& 512 & BN-I & - & 65.7 & 77.0 & 86.3 & 91.2 & 95.0 & 97.3 & - & 84.1 & 90.4 & 94.0 & 96.5 & 98.0 & 98.9 \\
			\multirow{1}{*}{VML~\cite{kim2021virtual}} 
			& 512 & BN-I & - & 66.4 & 76.9 & 86.7 & 91.3 & - & - & - & 84.6 & 91.1 & 95.1 & 97.2 & - & - \\
			\multirow{1}{*}{CircleLoss~\cite{Authors61}} 		
			& 512 & BN-I & - & 66.7 & 77.4 & 86.2 & 91.2 & - & - & - & 83.4 & 89.8 & 94.1 & 96.5 & - & - \\
			\multirow{1}{*}{RMS~\cite{wang2021ranked}}
			& 512 & BN-I & - & 67.4 & 78.4 & 86.6 & 91.8 & 95.5 & 97.9 & - & 84.6 & 90.8 & 94.3 & 96.8 & 98.5 & 99.2 \\
			
			\hline
			
			\multirow{1}{*}{\textbf{DDTAS}} 			
			& 512 & BN-I & \textbf{71.0} & \textbf{68.4} & \textbf{78.7} & \textbf{86.7} & \textbf{92.1} & \textbf{95.6} & \textbf{97.7} & \textbf{73.3} & \textbf{86.4} & \textbf{92.0}  &\textbf{95.4}  &\textbf{97.2}  & \textbf{98.5} & \textbf{99.2}\\
			\hline
	\end{tabular}}	
\end{table*}

Combining (a) and (c) in Fig. \ref{pncmp}, we can see that for deep metric learning, in the specific training process, the number of positive pairs has been insufficient for a long time. As seen from the first column in each subfigure, the algorithm with the same threshold can allow as many positive samples to participate in training as possible. But in return, the negative pairs have a more significant number throughout the training phase. The abundance of negative pairs introduces excessive redundant information, thereby disrupting the compactness of the distribution of negative samples no longer compact (column 1 in Fig. \ref{pncmp} (d)). This directly leads to the degradation of image retrieval performance. When we chose the second threshold screening strategy, the number of negative pairs dropped significantly, the model learned more valuable negative pair information and improved image retrieval performance. But the disadvantage of shared thresholds is shown again here. Positive pairs cannot exert total training value due to too strict filtering conditions. The over-compact similarity distribution (column 2 in Fig. \ref{pncmp} (d)) also makes it difficult to separate the positive and negative sample pairs. 

Our ASMS imposes strong screening conditions on negative pairs, while positive pairs are selected with mild tolerance (column 3 in Fig. \ref{pncmp} (a)-(d)). This strategy can ensure that more informative pairs are mined, enabling a sufficient number of positive sample pairs to participate in training. And through (b), we can find an interesting phenomenon. After we use our ASMS, in the middle and late stage of training, the selected sample pairs appear many uncommon sample size states with more positive pairs and few negative sample pairs. This is beneficial for model training. At the same time, from the similarity distribution of positive and negative pairs (column3 in Fig. \ref{pncmp} (d)), this strategy allows the model to learn superior embedding space.

\subsection{Comparison With Existing Algorithms}
In this section, we compare the performance of our algorithm with other state-of-the-art approaches on image retrieval tasks. As shown in Tables~\ref{CMP1} and \ref{CMP2}, our algorithm achieves the best performance on both the CUB200 and Cars196 datasets. The performance has been improved by 1\% (67.4\% $\rightarrow$ 68.4\%) and 1.8\% (85.2\% $\rightarrow$ 86.4\%) on Recall@K1, respectively, compared with the second-ranked RMS~\cite{wang2021ranked} and ABE~\cite{Authors59}. Compared with MS Loss, which also utilizes similarity and tolerance thresholds for sample selection, our results on CUB200 and Cars196 improved by 2.7\% (65.7\% $\rightarrow$ 68.4\%) and 2.3\% (84.1\% $\rightarrow$ 86.4\%), respectively. On the larger-scale dataset SOP, our algorithm is still competitive. It needs to be mentioned here that too many categories of SOP will make it impossible to read all the category data of the meta-learning set in a single iteration when using the threshold generator. Therefore, when using the threshold generator on SOP, we replace the single-step gradient descent on CUB200 and Cars196 by training an entire epoch of data. Our algorithm achieves 78.0\% for image retrieval performance, which is also an excellent result.

\section{Discussion}\label{secd}

\textbf{Limitation.} Our proposed DDTAS utilizes an adaptive threshold generator inspired by meta-learning to reduce the tedious manual parameter tuning process in existing methods. This module requires executing a training task separate from the main training process to generate new thresholds. Although we employ a lightweight meta-learning set to minimize resource consumption, the generator inevitably increases computational costs.\\
\textbf{Impact and future work.} Through dynamic threshold adjustments for adaptive pair mining, our method significantly enhances performance. The achievements obtained by our method reveal the potential of devising dynamic and flexible pair selection strategies, which can inspire researchers to establish more muscular deep metric learning models.
In future work, we intend to delve into parameter-tuning models with low computational costs to further enhance the efficiency of our method. Additionally, we plan to evaluate our strategy on more diverse datasets to thoroughly validate the robustness of our approach in addressing challenges such as varying lighting conditions, noise levels, image clarity, and data scale. Furthermore, we aim to extend the applicability of our method beyond image retrieval tasks to broaden the scope of practical utility for our strategy.

\begin{table}[t]
	\begin{center}
		\renewcommand{\arraystretch}{1}
		\caption{Recall@$K$ (\%) performance on SOP.}
		\label{CMP2}
		\resizebox{0.5\textwidth}{!}{\begin{tabular}{l|c|c|cccc}
				\hline
				\multicolumn{1}{c|}{\multirow{2}{*}{\textbf{Method}}}& \multicolumn{1}{c|}{\multirow{2}{*}{\textbf{Dim}}}  & \multicolumn{1}{c|}{\multirow{2}{*}{\textbf{BB}}} & \multicolumn{4}{c}{\textbf{SOP}} \\
				\ & & & 1 & 10 & 100 & 1000 \\	
				\hline
				\multirow{1}{*}{LiftedStruct~\cite{Authors49} }
				& 64 & G & 62.1 & 79.8 & 91.3 & 97.4\\
				\multirow{1}{*}{HDC ~\cite{Authors57}}
				& 384 & G & 69.5 & 84.4 & 92.8 & 97.7\\
				\multirow{1}{*}{BIER ~\cite{Authors58}}
				& 512 & G & 72.7 & 86.5 & 94.0 & 98.0 \\
				
				\multirow{1}{*}{A-BIER~\cite{Authors22}}
				& 512 & G & 74.2 & 86.9 & 94.0 & 97.8\\
				\multirow{1}{*}{ABE~\cite{Authors59}} 			
				& 512 & G & 76.3 & 88.4 & 94.8 & 98.2\\
				
				\multirow{1}{*}{LoOp~\cite{vasudeva2021loop}}
				& 512 & G & 76.6 & 89.8 & 95.8 & -\\
				\multirow{1}{*}{XBM~\cite{Authors50}}
				& 512 & G & 77.4 & 89.6 & 95.4 & 98.4\\ 
				\hline
				\multirow{1}{*}{Margin~\cite{Authors44}}
				& 128 & R-50 & 72.7 & 86.2 & 93.8 & 98.0\\
				\multirow{1}{*}{PADS~\cite{Authors68}} 			
				& 512 & R-50 & 76.5 & 89.0 & 95.4 & -\\
				\multirow{1}{*}{MIC~\cite{Authors44}}
				& 128 & R-50 & 77.2 & 89.4 & 95.6 & -\\
				
				\hline
				\multirow{1}{*}{Clustering ~\cite{Authors54}}
				& 64 & BN-I & 67.0 & 83.7 & 93.2 & - \\
				\multirow{1}{*}{\textit{N}-Pair-Loss\cite{Authors56} }
				& 64 & BN-I & 67.7 & 83.8 & 93.0 & 97.8\\
				\multirow{1}{*}{ProxyNCA~\cite{Authors55}} 
				& 64 & BN-I & 73.7 & - & - & -\\
				\multirow{1}{*}{ALA~\cite{Authors62}}
				& 512 & BN-I & 77.0 & 89.4 & 96.1 & -\\
				\multirow{1}{*}{VML~\cite{kim2021virtual}} 
				& 512 & BN-I & 77.9 & 90.3 & 96.0 & -\\
				\multirow{1}{*}{RMS~\cite{wang2021ranked}}
				& 512 & BN-I & 78.1 & 90.9 & 96.9 & -\\
				\multirow{1}{*}{MS~\cite{Authors7}}
				& 512 & BN-I & 78.2 & 90.5 & 96.0 & 98.7\\
				\multirow{1}{*}{SoftTriple~\cite{Authors60}}
				& 512 & BN-I & 78.3 & 90.3 & 95.9 & -\\
				\multirow{1}{*}{CircleLoss~\cite{Authors61}}
				& 512 & BN-I & 78.3 & 90.5 & 96.1 & 98.6\\
				
				\hline
				\multirow{1}{*}{\textbf{DDTAS}} 		
				& 512 & BN-I &  \textbf{78.0} &  \textbf{90.4} & \textbf{96.0} & \textbf{98.3}\\
				\hline
		\end{tabular}}		
	\end{center}
\end{table}

\section{Conclusion}\label{sec6}

This work focuses on thresholds in sample mining strategies and loss functions within the domain of deep metric learning. We propose differentiated dynamic adjustment strategies for the thresholds in these two key aspects, collectively terms the Dual Dynamic Threshold Adjustment Strategy (DDTAS). Within this framework, we propose an Asymmetric Sample Mining Strategy (ASMS) tailored for sample mining to address the challenges related to imbalanced sample pairs and a deficiency of positive pairs in the context of deep metric learning. Then ASMS is further optimized into the Adaptive Tolerance Asymmetric Sample Mining Strategy (AT-ASMS), which can flexibly adapt the threshold according to the ratio of positive and negative samples. Extensive experiments have shown that our proposed method has achieved the state-of-the-art performance in image retrieval.

\bibliographystyle{IEEEtran}
\bibliography{references}

\end{document}